\documentclass[prl,twocolumn,english,superscriptaddress]{revtex4-1}
\usepackage{graphicx}
\usepackage{color}
\usepackage{epstopdf,gensymb}
\usepackage{amsmath,amsthm,amsfonts,amscd}
\begin{document}

\title{A Relativistic Plasma Polarizer: Impact of Temperature Anisotropy on Relativistic Transparency}
\author{David J. Stark}
\author{Chinmoy Bhattacharjee}
\affiliation{Department of Physics, The University of Texas at Austin, Texas 78712, USA.}
\affiliation{Institute for Fusion Studies, The University of Texas at Austin, Texas 78712, USA.}
\author{Alexey V. Arefiev}
\affiliation{Institute for Fusion Studies, The University of Texas at Austin, Texas 78712, USA.}
\author{R. D. Hazeltine}
\author{S. M. Mahajan}
\affiliation{Department of Physics, The University of Texas at Austin, Texas 78712, USA.}
\affiliation{Institute for Fusion Studies, The University of Texas at Austin, Texas 78712, USA.}

\date{\today}
\pacs{52.35.-g; 52.35.Hr; 52.35.We; 67.10.-j; 67.30.hj}

\begin{abstract}

3D particle-in-cell simulations demonstrate that the enhanced transparency of a relativistically hot plasma is sensitive to how the energy is partitioned between different degrees of freedom. For an anisotropic electron distribution, propagation characteristics, like the critical density, will depend on the polarization of the electromagnetic wave. Despite the onset of the Weibel instability in such plasmas, the anisotropy can persist long enough to affect laser propagation. This plasma can then function as a polarizer or a waveplate to dramatically alter the pulse polarization.

\end{abstract}

\maketitle
When the electron population in a plasma reaches relativistic energies, the dielectric properties can change drastically enough for the plasma to become transparent to an electromagnetic (EM) wave that cannot penetrate a low-energy plasma of the same density. When the electrons are brought to these energies directly by the electromagnetic pulse (of high intensity), the resulting phenomenon is called self-induced transparency~\cite{Palaniyappan_2012}. The enhanced transparency, however, is an intrinsic  characteristic of a relativistically hot plasma independent of the source of heating. In the era of high-power lasers, when experimental studies of relativistic plasmas are possible for a staggering variety of applications (proton therapy, material studies, laboratory astrophysics, basic dynamics), a detailed understanding of relativistic transparency will be essential, both for a proper interpretation of experiments and as a new diagnostic tool.  Earlier theoretical studies of self-induced transparency dealt with high amplitude propagating solutions in homogeneous and weakly inhomogeneous plasmas~\cite{Akhiezer_1956, Kaw_1970, Max_1971, Marburger_1975, Lai}. Most recently, progress has been made in understanding the plasma-wave interaction at the plasma-vacuum interface and the onset of relativistic transparency as a high intensity pulse irradiates a cold plasma slab~\cite{Cattani,Goloviznin,Tushentsov,Vshikov,Bulanov,Weng2, Eremin_2010, Siminos_2012}.

In most studies on the subject, focused on determining how the transparency threshold scales with both the plasma density and the intensity of the irradiating pulse, the pulse serves the dual purpose of imparting relativistic energy  to electrons and simultaneously acting as a probe of criticality. These experiments, concentrating on the total electron energy, do not fully investigate the role that the shape of the electron distribution could play in determining the transparency threshold. The approach is consistent with the commonly used explanation that the relativistic mass increase, by lowering the plasma frequency, raises the critical density below which  the electromagnetic waves  are able to propagate. Since the relativistic $\gamma$-factor is a gross  measure of the overall energy, this explanation could not reveal if the propagation characteristics are affected by the way the energy is partitioned between different degrees of freedom. Because the critical density for electromagnetic waves in warm non-relativistic plasmas is independent of the shape of the electron distribution, a similar conclusion in the relativistic case may appear to be justifiable; most experiments are designed and interpreted within this context.

However, one could envision an alternative system setup in which a plasma is heated to relativistic temperatures by a high-power pump pulse and then probed with a low-amplitude pulse, allowing the properties of the created distribution function to be tested without changing the distribution itself. Indeed, several experiments have used a transverse optical probe pulse to hit the system during the laser-plasma interaction as a means of measuring specific properties of the system~\cite{Buck,Schwab,Jackel, Kaluza}. Characterizing relativistic transparency's effects on pulse propagation enables the probe to serve as a diagnostic for the plasma energy, temperature, and especially anisotropy. This information, in turn, is crucial to the interpretation and prediction of the high-amplitude pulse's behavior in the plasma. The better characterization of a laser-produced distribution has particular relevance to laboratory astrophysics~\cite{Bulanov2, Esirkepov} and ion acceleration from laser-irradiated solid-density targets~\cite{Yin_2011, Shorokhov_2004,Hegelich_2013,Sahai_2013,Macchi_2013}.

In this Letter we demonstrate that relativistic transparency is strongly affected by  how the electron energy is partitioned between different degrees of freedom. We consider here the simplest problem: the propagation of a low amplitude pulse through  a preformed relativistically hot anisotropic electron plasma (ion motion is neglected) to explore its intrinsic dielectric properties  (unchanged by the weak pulse). We find that:  1) the critical density for propagation depends strongly on the pulse polarization, 2) two plasmas with the same density and average energy per electron can exhibit profoundly different responses to electromagnetic pulses, 3) the anisotropy-driven Weibel instability develops as expected; the timescales of the growth and back reaction (on anisotropy), however, are long enough that sufficient anisotropy persists for the entire duration of the simulation, consequently impacting the optical properties. Modified propagation characteristics add a qualitative new element in developing  a more advanced understanding of laser-plasma interactions and even broader astrophysical phenomena. These effects will likely serve as the foundation for developing new optical devices utilizing the capability of manipulating pulse polarizations~\cite{Michel}.

Using a 3D-3V particle-in-cell simulation (three spatial and three velocity dimensions), we study the dynamics of a low amplitude circularly polarized electromagnetic pulse incident on a finite slab of constant density electrons (ions fixed) with an anisotropic relativistic temperature. The domain is $130\mu$m$\times 70\mu$m$\times 70\mu$m ($4500\times100\times100$ cells) and consists of vacuum regions at $-30\mu m<x<0$ and $8\mu m<x < 100\mu$m and a plasma region at $0<x<8\mu$m. A circularly polarized Gaussian pulse (full width half maximum (FWHM) of 50 fs) of wavelength $\lambda = 2$ $\mu$m enters the plasma from negative $x$ and focuses halfway into the target with intensity FWHM of $11.8\mu$m. The pulse has focal amplitude of $a=|e|E_{0}/m_ec\omega=0.2$, where $E_{0}$ is the electric field amplitude, $\omega$ is the wave frequency, $c$ is the speed of light, and $m_e$ and $e$ are the electron mass and charge, respectively. The electron number density $n$ ramps up and falls off as a semi-Gaussian of FWHM $2.5\mu$m, so that $n=2.7n_{*}$ for $2.3\mu {\textrm{m}}< x < 5.7$ $\mu$m. Here $n_{*} \equiv m_e \omega^2/{4\pi e^2}$ is the classical critical density. We use 120 electrons per cell to initialize 
an anisotropic momentum distribution given by  
\begin{equation}
\label{AnMaxdist}
\mathcal{\mathit{f}}_{0}=\frac{n}{I(\alpha,\beta)}\, \exp\left(-\alpha \sqrt{1+\frac{p_z^2+\beta(p_x^2+p_y^2)}{m_e^2 c^2}}\right),
\end{equation}
where $1/\alpha$ is an effective temperature normalized to $m_e c^2$ and $\beta$ introduces anisotropy into the distribution (when $\beta \neq 1$). In Eq. (\ref{AnMaxdist}), $n$ is the electron density, $p$ is the electron momentum, and $I$ is a dimensionless normalization constant
\begin{equation}
I(\alpha,\beta) = \int \exp \left(-\alpha \sqrt{1+\frac{p_z^2+\beta(p_x^2+p_y^2)}{m_e^2 c^2}} \right) \frac{d^3p}{(m_e c)^3}.
\end{equation}
For $\beta < 1$,  the motion along the $z$-axis is always associated with less energy than in the other directions. In the simulation we use $\alpha=2.0$ and $\beta=0.55$, corresponding with average particle energy $<E>=1.24$ MeV, and $\sqrt{<p_y^2>/<p_z^2>}=1.35$, the latter being a measure of anisotropy. Here the brackets represent an average over the entire momentum space so that $<R> \equiv \int R f_{0} {d^3p}/{(m_e c)^3}$.

\begin{figure}[t]
	\centering
	\includegraphics[width=0.45\textwidth]{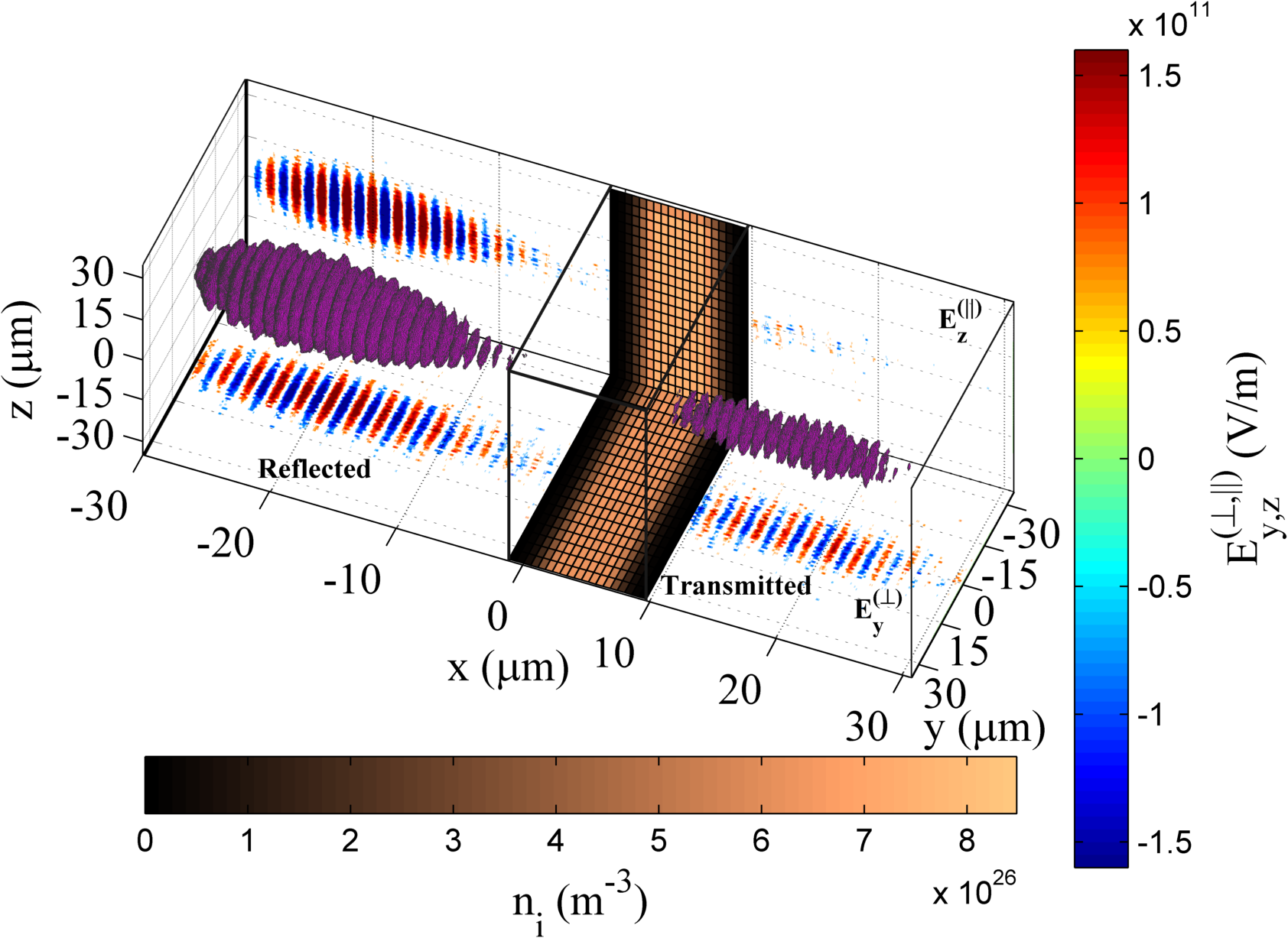}
	\includegraphics[width=0.35\textwidth]{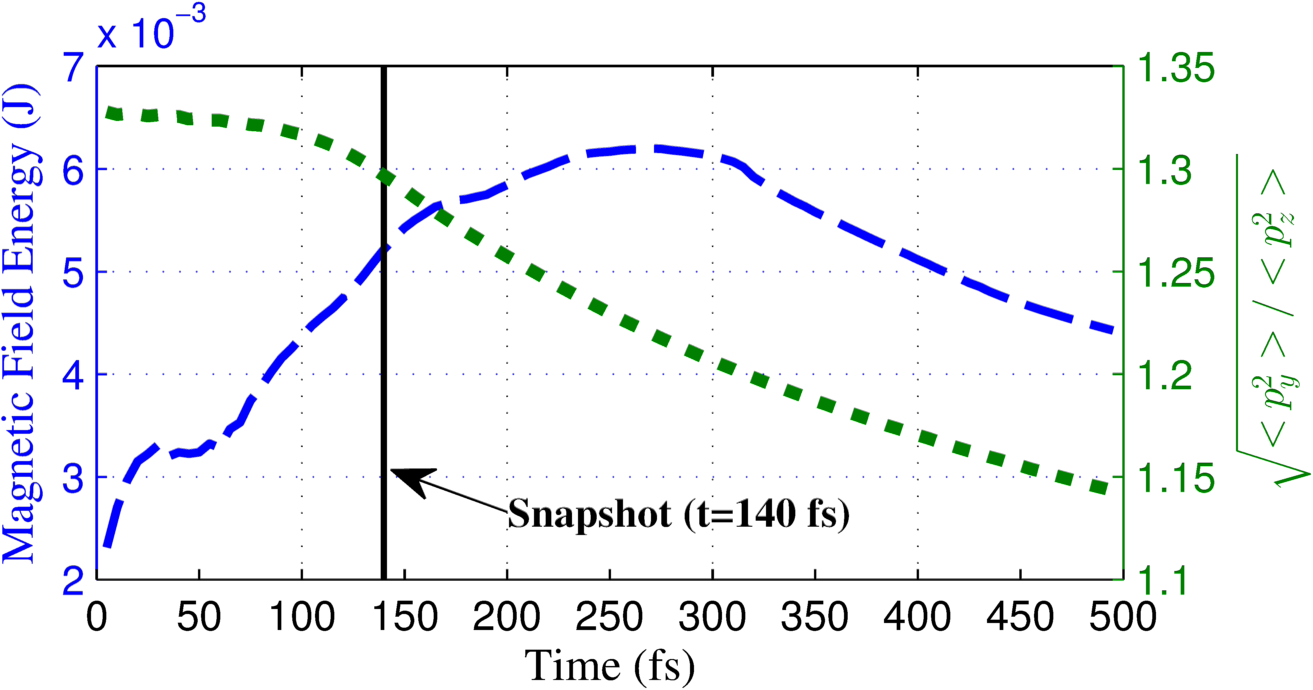}
	\caption{Surfaces of constant intensity of the reflected and transmitted pulses 140 fs after the pulse hits the target (top panel). $E_y$ and $E_z$ cross-sections at $z=0$ are given on the bottom and side of the box, respectively, along with $n_i$. The magnetic field energy and $\sqrt{<p_y^2>/<p_z^2>}$ are plotted throughout the simulation time (bottom panel).}
	\label{Polarizations}
\end{figure}

The simulation begins ($t=0$) with the leading edge of the circularly polarized pulse at $x=0$. In Figure~\ref{Polarizations}, the transmitted and reflected pulses are shown $140$ fs into the simulation. The pink surfaces denote surfaces of constant $E^2=8.0\times 10^{22}$ $($V/m$)^2$, whereas the images on the bottom and side of the box represent $E_y$ and $E_z$ at $z=0$. The ion number density is also projected onto the bottom and side of the box to show where the plasma resides. The simulation results are quite spectacular: the plasma acts as a powerful polarizer; it reflects almost all of  the parallel component (to the axis of anisotropy $z$), $E_z\equiv E_{\parallel}$, while it transmits much of the perpendicularly polarized component, $E_y\equiv E_{\perp}$. The latter hotter direction is favored for propagation.
%

Since an anisotropic electron distribution is subject to the Weibel instability \cite{weibel}, we have carefully monitored the growth of energy stored in the magnetic field of the system ($\int B^2/8\pi dV$). We have  displayed in  Figure~\ref{Polarizations} both the magnetic energy and the anisotropy parameter $\sqrt{<p_y^2>/<p_z^2>}$ as  functions of the simulation time.  Note that $\sqrt{<p_y^2>/<p_z^2>}$ starts slightly below the analytically predicted value of 1.35 at $t=0$, a discrepancy of order $2\%$. This can be improved by increasing the sampling resolution of the distribution function (shrinking the $p_x$, $p_y$, and $p_z$ step-size). Recent work has shown  kinetic simulations of the relativistic Weibel instability from thermal anisotropy~\cite{Kaang,Ghizzo}, which also exhibit, similar to our results, a peak in magnetic field energy right before falling to an asymptotic value. Here we observe that the anisotropy persists in the plasma over a sufficiently long timescale for the optical properties of this distribution to be probed. The pulse has already passed through the plasma well before $\sqrt{<p_y^2>/<p_z^2>}$ has appreciably diminished.


We next calculate, analytically, the critical frequency and density for the plasma distribution invoked in the simulation [see Eq. (\ref{AnMaxdist})]. A simple linear analysis for wave propagation will  demonstrate the disparity in critical densities based on polarization. Some examples of earlier studies of anisotropic plasmas  are~\cite{Schlickeiser1,Schlickeiser2,Schlickeiser3}. The basic dynamics is contained in the covariant Vlasov and Maxwell's equations (the momentum four-vector $p^{\mu}$ is normalized to $m_e$, and $c=1$):
\begin{align}
 \left[p^{\mu}\partial_{\mu}+qp_{\nu}F^{\mu\nu}\frac{\partial}{\partial p^{\mu}}\right]f(x,p)&=0, \label{vlasov} \\ 
 \partial_{\mu}F^{\mu\nu}&=4\pi J^{\nu}, \label{maxwell}
\end{align}
where $f(x,p)$ is the electron distribution function, $J^{\nu}=q/m\int d^{4}pp^{\nu}f(x,p)$ is the four-current, and $F^{\mu\nu}=\partial^{\mu}A^{\nu}-\partial^{\nu}A^{\mu}$ is the electromagnetic field tensor, $A^{\mu}$ being the potential four-vector. The summation convention is used, with metric $(+,-,-,-)$. We linearize Eqs.~(\ref{vlasov}) and (\ref{maxwell}), and assume perturbations of the form $f_{1},F_{1},A_{1}\propto\exp(- ik_{\mu}x^{\mu})$, choosing $k^{\mu}=(\omega,k,0,0)$.
In a field-free plasma, and  for the equilibrium distribution given by Eq. (\ref{AnMaxdist}), the two transverse modes $A_1^y$ and $A_1^z$ are decoupled, each producing current only parallel to its respective polarization. From these independent dispersion relations, the expression of the critical frequency for each mode is derived by setting $k=0$ and solving for $\omega$:
\begin{align} 
 \label{crit_cold}\omega_{\perp\mbox{c}}^{2} &= \frac{\alpha\beta\omega_{p0}^2}{I(\alpha,\beta)}\int d^3p \frac{p_y^2\exp \left(-\alpha \sqrt{1+p_z^2+\beta(p_x^2+p_y^2)} \right)}{\sqrt{1+p^2}\sqrt{1+p_z^2+\beta(p_x^2+p_y^2)}} ,\\
\label{crit_hot} \omega_{||\mbox{c}}^{2} &= \frac{\alpha\omega_{p0}^2}{I(\alpha,\beta)}\int d^3p \frac{p_z^2\exp \left(-\alpha \sqrt{1+p_z^2+\beta(p_x^2+p_y^2)} \right)}{\sqrt{1+p^2}\sqrt{1+p_z^2+\beta(p_x^2+p_y^2)}} ,
\end{align}
where $\omega_{p0} \equiv \sqrt{4 \pi n e^2/m_e}$ is the plasma frequency. The subscripts ${\perp}$ (${||}$) for  the modes with nonzero $A_y$ ($A_z$) indicate the  direction of  the electric field in relation to the  axis of anisotropy. One can readily find the critical densities for each mode directly from Eqs.~(\ref{crit_cold}) and (\ref{crit_hot}): $n_{||,\perp\mbox{c}}=(\omega_{p0}/\omega_{||,\perp\mbox{c}})^2n_{*}$, again with $n_{*} \equiv m_e \omega^2/{4\pi e^2}$. 

The expressions for $\omega_{\perp\mbox{c}}$ and $\omega_{||\mbox{c}}$ become more tractable in the case of weak anisotropy, i.e,  for $\beta$ close to one ($\beta=1-\epsilon$ where $\epsilon << 1$). To the leading order, 
\begin{align}
\label{cold_approx}\omega_{\perp\mbox{c}}^{ 2} &= \frac{\alpha^2 \omega_{p0}^2}{K_2(\alpha)} \Bigl[ \int_{\alpha}^{\infty} dz \frac{K_2(z)}{z^2}-2\epsilon G(\alpha)\Bigr] \\
\label{hot_approx}\omega_{||\mbox{c}}^{2} &= \frac{\alpha^2 \omega_{p0}^2}{K_2(\alpha)} \Bigl[ \int_{\alpha}^{\infty} dz \frac{K_2(z)}{z^2}-\epsilon G(\alpha)\Bigr],\\
\label{GG} G(\alpha) &= \int_{\alpha}^{\infty} dz \frac{K_2(z)}{z^2}+\frac{1}{2}\int_{\alpha}^{\infty} dz \left(\frac{\alpha^2}{z^3}-\frac{1}{z}\right)K_3(z)
\end{align}
where $K_i$ is the modified Bessel function of the second kind of order $i$. For the isotropic distribution ($\epsilon = 0$), naturally $n_{||\mbox{c}}= n_{\perp\mbox{c}}$. Notice that $n_{\perp\mbox{c}} > n_{||\mbox{c}}$ for $\epsilon > 0$; the critical density is lower for a wave whose electric field is polarized along the axis of anisotropy, that is, the colder direction in this simulation. This result is consistent with the presented simulation, where the density was $n=2.70n_{*}$. For the simulation's laser frequency,  $n_{\perp\mbox{c}}=2.74n_{*}$ and $n_{||\mbox{c}}=2.50n_{*}$, so that the $y$-component should pass through, whereas the $z$-component should not.


\begin{figure}[t] 
	\centering
    \includegraphics[width=0.37\textwidth]{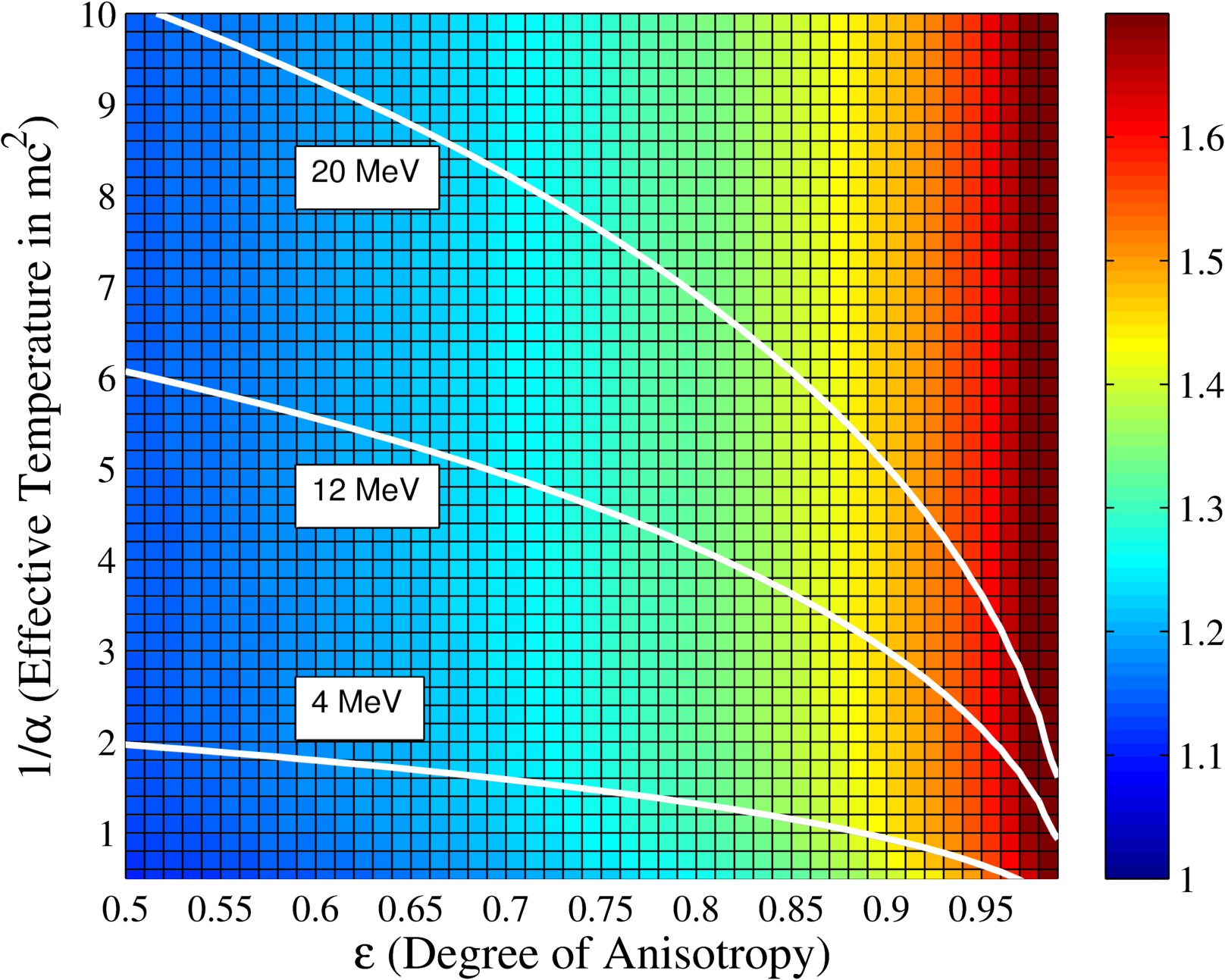}
	\caption{Ratio of critical densities, $n_{\perp\mbox{c}}/n_{||\mbox{c}}$, as a function of effective electron temperature $1/\alpha$ and the degree of anisotropy $\epsilon$. The solid lines indicate the contours of constant average electron energy.}
	\label{analytical_comparison}
\end{figure}

In the non-relativistic limit  $({\alpha}^{-1}\sim T_e/m_e\ll1)$, the anisotropic terms in Eqs. (\ref{cold_approx}) and (\ref{hot_approx}) become vanishingly small: $\alpha^2G(\alpha)/K_2(\alpha)\to \mathcal{O}({\alpha}^{-1})$; the critical densities, then, show no discernible difference between the two polarizations. In this sense, it is fundamentally the relativistic effects that can cause strong differences in transparency between the polarizations.

Figure~\ref{analytical_comparison} shows how the disparity in critical densities between the two polarizations, calculated from Eqs. (\ref{crit_cold}) and (\ref{crit_hot}), increases both with the increase of electron energy and with the degree of anisotropy $\epsilon$ (here $\epsilon$ is not necessarily small). The solid lines indicate the contours of constant $<E>$, and the ratio $n_{\perp\mbox{c}}/n_{||\mbox{c}}$ changes considerably along these contours; the relativistic transparency varies rapidly with anisotropy in the electron distribution even when the average energy is kept constant.


The anisotropy-induced discrepancy in the critical densities has a profound effect on wave propagation; this is true even when 
the plasma is transparent (low density) to arbitrary polarization. For demonstration, we conduct a 3D-3V simulation, complementary to the earlier one, in which a linearly polarized Gaussian pulse is incident on a finite length sub-critical plasma with an anisotropic distribution. The same electron distribution is used, but now
the density  $n=1.75 n_{*}$ is chosen so that both $n<n_{\perp\mbox{c}}$ and $n<n_{||\mbox{c}}$. The pulse is polarized at a 45 degree angle to the axis of anisotropy, so that $E_y = E_z$ in the incoming pulse. The pulse width has FWHM $50$ fs, peak $a=0.4$, $\lambda =2.0$ $\mu$m, and intensity FWHM $11.8\mu$m. The simulation domain now consists of a vacuum region at $x < 0$ and a plasma region at $0\leq x \leq 25\mu$m. The density ramps up and falls off as a semi-Gaussian of FWHM $3.3\mu$m, so that $n=1.75n_{*}$ for $4.3\mu {\textrm{m}}< x < 20.7$ $\mu$m. 

\begin{figure}[t]
	\centering
    \includegraphics[width=0.5\textwidth]{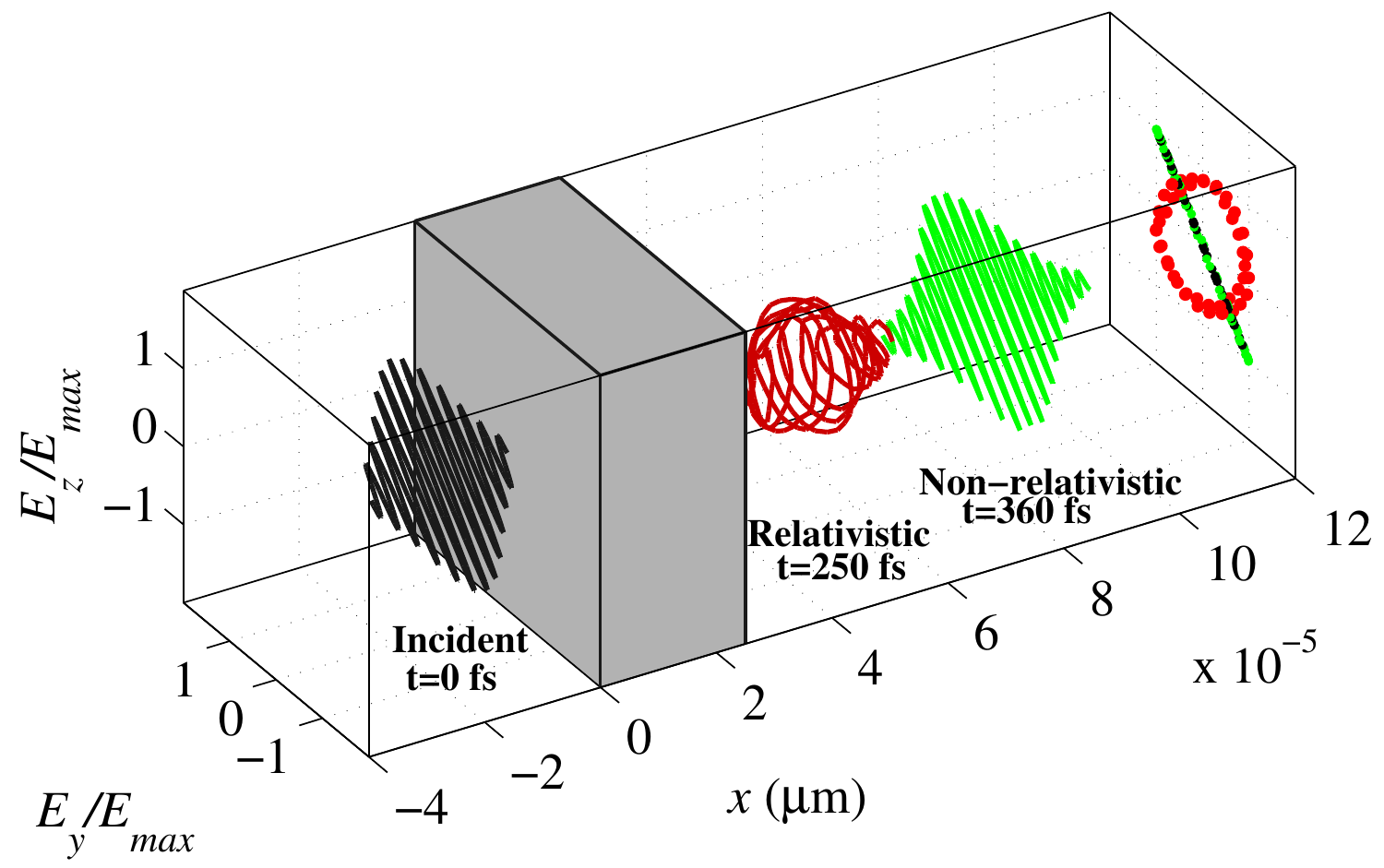}
  
	\caption{Plots of the incident pulse (at $t=0$), the transmitted pulse through the relativistic plasma (at $t=250$ fs, $\alpha=2.0$ and $\beta=0.55$), and the transmitted pulse through the non-relativistic plasma (at $t=360$ fs, $\alpha=500$ and $\beta=0.55$) are denoted in black, red, and green respectively. }
	\label{figlast}
\end{figure}

The incoming pulse can be decomposed into two modes: one polarized along the axis of anisotropy and the other perpendicular to it. These modes are decoupled and have two different critical densities.  
For a cold plasma, the group velocity $v_g$ for $0 < (\omega - \omega_{p0})/\omega_{p0} << 1$ scales as $1/\omega_{p0}$, where $\omega_{p0}$ (the critical frequency in this case) is determined by the critical density. Therefore,  due to the difference between $n_{||\mbox{c}}$ and $n_{\perp\mbox{c}}$, we expect a considerable discrepancy in group velocities between the two modes. In Figure~\ref{figlast}, following the pulse before (black) and after (red) it passes though the plasma, we see that the induced spatial separation of the two modes changes the pulse from linear to elliptical polarization, highlighting the expected discrepancy in $v_g$; in this scenario the plasma serves as a waveplate. Both of our simulations demonstrate how a relativistic plasma can change the polarization of an electromagnetic wave; naturally the excess (shortage) of the wave angular momentum is compensated by the corresponding loss (gain) by the plasma.


Figure~\ref{figlast} clearly demonstrates that  anisotropy-induced polarization change is essentially a relativistic phenomenon. In a non-relativistic anisotropic plasma the group velocity depends on the polarization, but the critical density does not. Consequently, the linear polarization of the wave remains  essentially unaffected after the initial pulse (black) propagates through the plasma (green). The parameters for the non-relativistic simulation are: the anisotropy index $\sqrt{<p_y^2>/<p_z^2>}=1.35$, $\alpha=500$, and $\beta=0.55$ with an average kinetic energy $\approx 0.002$ MeV, much smaller than the rest mass energy. We also set $n=0.7n_{*}$ to ensure that both polarizations penetrate the plasma and reduce the amplitude to $a=0.1$. The relativistic anisotropic plasma, in stark contrast, changes the linear polarization to elliptical.


An investigation of the interaction of electromagnetic waves with relativistically anisotropic plasmas, thus, reveals a new qualitative phenomenon: the propagation characteristics  (critical density, effective refractive index) of the wave are controlled  not only by plasma density and average electron energy, but also by how the energy is partitioned between different degrees of freedom, i.e, by anisotropy. An anisotropic plasma emerges as an effective  polarizer; it will filter out the electric field of the pulse polarized in the ``colder" direction, and pulses of the same frequency, polarized in the hot direction, will be preferentially transmitted. Even if the plasma is transparent for all polarizations, the discrepancy in the critical densities causes spatial separation of the modes, manifested as an altered polarization of the pulse so that the plasma here serves as a waveplate. Since, for a given anisotropy,  it is only at  relativistic energies that differential propagation becomes evident,  high power laser-plasma experiments must take account of this effect for proper interpretation. This is particularly relevant for interpreting the data from probe pulses simultaneously incident on the plasma with the pump pulse. Polarization shifts in the probe pulse over time serve as a measure of the temperature anisotropy in the created system, along with the evolution of the anisotropy on the timescale of the ultra-short pulse. These shifts could compete with Faraday rotation of the probe pulse used in magnetic field measurements~\cite{Hoth}. 
One could also envision utilizing anisotropic plasma as the basis for new optical devices used for beam polarization or polarization smoothing~\cite{Lefebvre,Munro,Rothenberg}.
While our parameters are relevant to these contemporary laser-plasma systems, our results are generic and equally apply to different parameter regimes. Differential propagation characteristics can affect high-harmonic and synchrotron transmission through dense laser-irradiated targets~\cite{Horlein,Dromey}. Finally, it is plausible that  such an effect could provide a possible explanation for polarization dependences  observed in  gamma-ray  bursts~\cite{Steele,Greiner,Coburn,Wiersema, Mundell}.

Simulations were performed using EPOCH code (developed under UK EPSRC grants EP/G054940/1, EP/G055165/1 and EP/G056803/1) using HPC resources provided by the TACC at The University of Texas. We acknowledge valuable discussions with Amir Shahmoradi, Patrick Crumley, and Toma Toncian. This work was supported by US DOE Contract No. DE-FG02-04ER54742, NNSA Contract No. DE-FC52-08NA28512, and DOE SCGF administered by ORISE-ORAU under Contract No. DE-AC05-06OR23100 (D. J. S.).

%

\end{document}